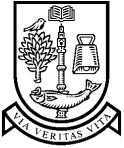

**UNIVERSITY** *of* **GLASGOW**

**Department of Physics & Astronomy**

Experimental Particle Physics Group
Kelvin Building, University of Glasgow,
Glasgow, G12 8QQ, Scotland.
Telephone: +44 - (0)141 – 3398855     Fax: +44 – (0)141 - 3305881

# Radiation Induced Damage in GaAs Particle Detectors.

R. Bates, C. Da Via, V. O'Shea, A. Pickford, C. Raine and K. Smith.
Department of Physics & Astronomy, University of Glasgow, Glasgow G12 8QQ, Scotland.

*Abstract*

*The motivation for investigating the use of GaAs as a material for detecting particles in experiments for High Energy Physics (HEP) arose from its perceived resistance to radiation damage. This is a vital requirement for detector materials that are to be used in experiments at future accelerators where the radiation environments would exclude all but the most radiation resistant of detector types.*

# Radiation Induced Damage in GaAs Particle Detectors.

R. Bates, C. Da Via, V. O'Shea, A. Pickford, C. Raine and K.Smith.
Department of Physics & Astronomy, University of Glasgow, Glasgow G12 8QQ, Scotland.

## Abstract

The motivation for investigating the use of GaAs as a material for detecting particles in experiments for High Energy Physics (HEP) arose from its perceived resistance to radiation damage. This is a vital requirement for detector materials that are to be used in experiments at future accelerators where the radiation environments would exclude all but the most radiation resistant of detector types.

## INTRODUCTION

GaAs particle detectors were developed for use in the experiments currently being designed as part of the new accelerator (the LHC) to be built at CERN, Geneva. The estimated radiation fluence foreseen for the innermost forward tracking layer of the experiment is of the order of several $10^{14}$ cm$^{-2}$ of particles, consisting of roughly equal fluxes of charged hadrons and neutrons with a typical energy of 1 MeV. This will be accumulated over the life of the experiment, which is projected to be 10 years.

GaAs had been proposed as a radiation hard detector material for tracking particles in this very forward region, where the radiation environment will be harshest and precludes the use of standard types of semiconductor radiation detectors. Extensive studies[1] have already been carried out on the resistance to radiation damange of this material using Gamma radiation. The purpose of this study was to evaluate the damage from neutrons and charged particles with particular emphasis on pions, as these are foreseen to be one of the dominant sources of damage to semiconductor particle detectors at the LHC.

## MEASUREMENTS AND RESULTS

All the detectors used in this study were 3 mm diameter circular Schottky diodes with a 500 μm wide guard ring separated by 10 μm on 200 μm thick semi-insulating GaAs substrates. The detector operates as a reverse biased diode, the energy deposited by a particle creating electron-hole pairs which are swept out of the depleted region by the applied field, which create a signal in the front-end charge amplifier. The main detector characteristics of interest for HEP are the leakage current, the full depletion voltage ($V_{fd}$) and the charge collection efficiency (cce). The leakage current is measured as an I-V characteristic up to the breakdown voltage or until the diode reaches a compliance current. $V_{fd}$ is rather more difficult to measure as there is incomplete charge collection in the first place. This value is measured by looking for alpha particle detection on the back contact as the range of this type of particle is approximately 20 μm in GaAs. This effect is also seen on the I-V curve as a knee in the leakage current corresponding to an increase in the injection current from the back contact.

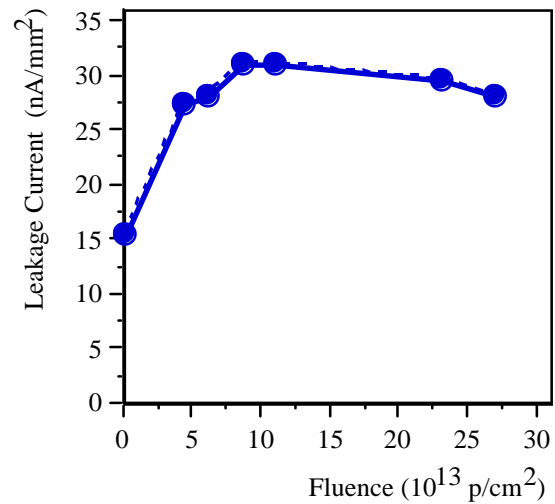

Fig[1] Detector leakage current increase vs. proton irradiation fluence. The current increases initially and then remains flat at a factor of 2 higher value in the range of interest.

The cce of the device is defined as the ratio of the measured charge to the total charge deposited in the detector by the ionising particle. The deposition of energy in the bulk of the detector depends on the type of radiation used to illuminate the detector. Beta particles with an energy of over 2 MeV deposit energy uniformly along their path through the detector material with a Landau energy distribution and are known as minimum ionising particles (MIPs). The most probable value of this deposited energy is 56 keV per 100 μm of GaAs traversed. Alpha particles lose all their energy close to the surface of the material because of their short range (approximately 20 μm for 5.5 MeV alphas from Am$^{241}$). This property is very useful as it permits the contribution to the total charge collected by the holes and the electrons to be separated by illuminating the detector from either side. This technique has been used to estimate the mean free path for each carrier in the material.

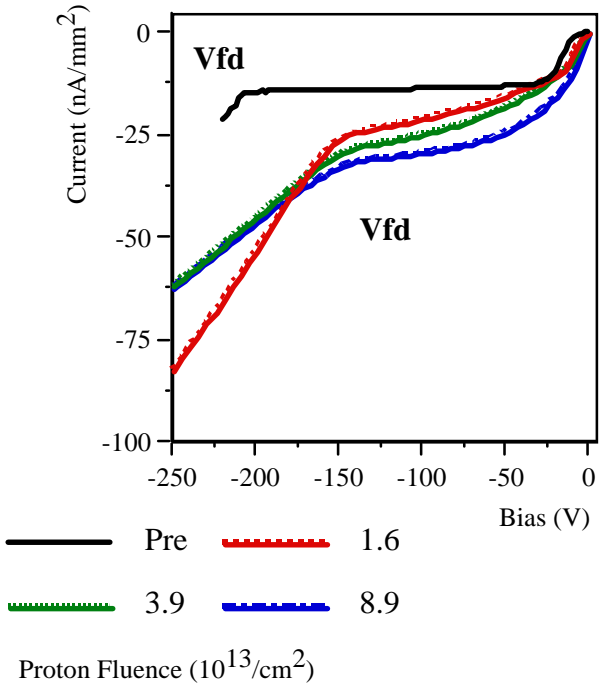

Fig[2] Leakage current for different fluences of protons. The change in $V_{fd}$ is indicated.

In previous studies devices had been irradiated up to 100 Mrad with $Co^{60}$ gammas and showed negligible degradation.

Neutrons from a spallation source at ISIS with a peak energy of 1 MeV, 24 GeV/c protons from the PS (Proton Synchrotron) at CERN and 300 MeV/c pions from the PSI facility at Villigen, Switzerland were used to irradiate the samples. The energy of the neutrons is the typical energy expected for the background neutron flux at the LHC.

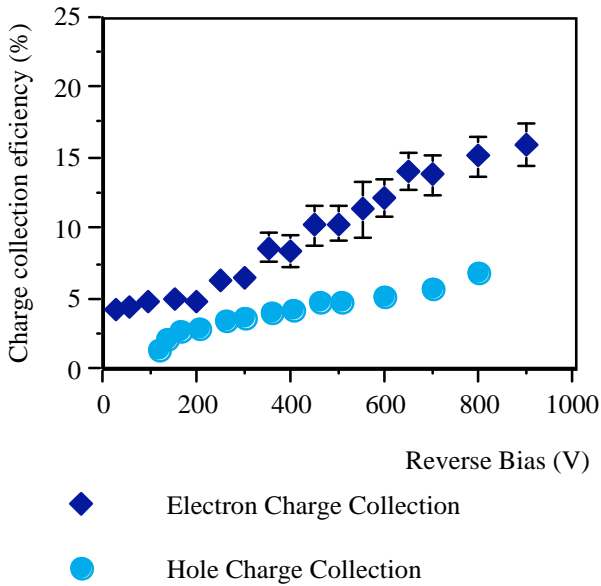

Fig[3] Electron and hole contributions to the charge signal after $2 \times 10^{14}$ p cm$^{-2}$ measured using alphas on the front and back of the device.

The pion energy is at the delta resonance and results in maximum damage. The proton energy is that used to irradiate silicon detectors for similar applications to those proposed for their GaAs counterparts.

The change in leakage current as a function of irradiation is of the order of a factor of two over the range of interest for all types of irradiation (Fig.[1]). This change is orders of magnitude lower than that for silicon detectors. This is not a cause for concern for the use of GaAs devices in the LHC environment, as the leakage current increase would contribute, at most, a 30% decrease in signal to noise ratio - an effect which is even further reduced by the fast shaping times proposed for the LHC experiments.

The voltage at which full depletion occurs ($V_{fd}$) decreases with increasing fluence for all types of charged particle irradiation (as illustrated in Fig.2 for protons) and is consistent with a decrease in free carrier concentration similar to the donor removal effect seen in silicon before type inversion. From an initial fluence of around $2 \times 10^{13}$ p cm$^{-2}$, $V_{fd}$ falls by about 35% from its pre-irradiation value which is always around 1 V μm$^{-1}$ of substrate thickness for the type of detectors used in this study. At higher fluences $V_{fd}$ remains almost unchanged.

The cce for MIPs is the parameter most affected by particle irradiation. The loss of charge is due to the creation of radiation induced trapping centres in the bulk. These either trap the charge created by the traversing particle or modify the electric field within the bulk of the detector so as to enhance the trapping efficiencies of the traps present.

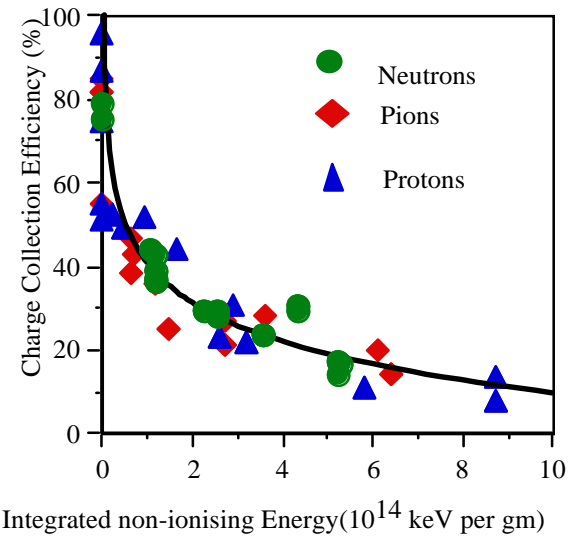

Fig[4] Charge collection efficiency as a function of integrated NIEL for protons, pions and neutrons. The energy deposited by different particles has been normalised according to Table 1.

The cce of a 200 μm thick detector, reverse biased at 200 V, before irradiation is typically >75% which corresponds to a signal greater than 20,000 electrons.

The reduction of cce with fluence depends on the type and energy of the irradiating particle. At a bias of 200 V a 10,000 electron signal is seen for a MIP (~40% cce) after $1.4 \times 10^{14}$ n cm$^{-2}$, $6.0 \times 10^{13}$ p cm$^{-2}$ and only $3.0 \times 10^{13}$ π cm$^{-2}$, for the energies used. Initially, the cce falls rapidly as a function of fluence and then decreases at a lower rate. The two stage cce dependence on fluence is similar to that observed in the dependence of leakage current as a function of fluence.

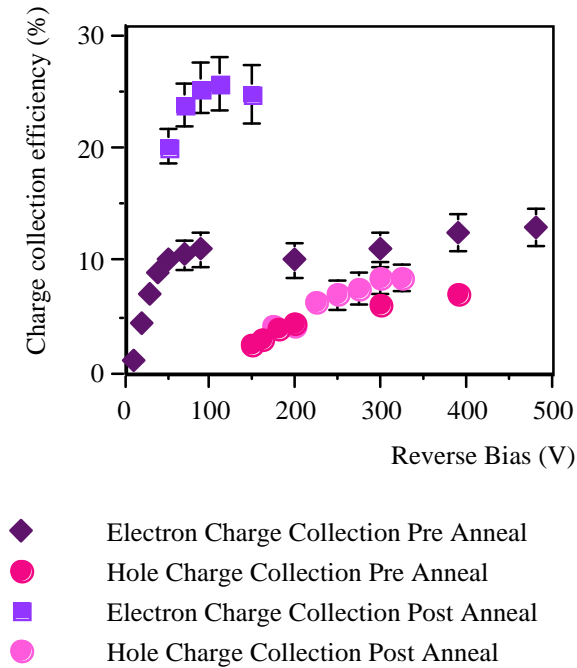

- ◆ Electron Charge Collection Pre Anneal
- ● Hole Charge Collection Pre Anneal
- ■ Electron Charge Collection Post Anneal
- ● Hole Charge Collection Post Anneal

Fig[5] The effect of a thermal anneal at 450 °C for 90 seconds on a diode that has been irradiated with $1 \times 10^{14}\ \pi\ cm^{-2}$.

From alpha cce data it has been shown that the signal from holes falls faster with increasing fluence than that from electrons and so the MIP signal after irradiation is mainly due to the electron signal. If the bias is increased the signal begins to increase again and this is due to a further increase in the electron signal as shown in Fig[3].

The differing effects from each type of irradiation on the cce can be attributed to the differing amounts of non-ionising energy loss (NIEL) as the particles pass through the crystal lattice. Fig[4] shows the correlation between cce for 200 μm thick detectors measured at 200 V and the calculated NIEL[2,3] for the various particles used in this study. The damage attributed to each particle type has been weighted by its relative NIEL and the calculated energy loss is well correlated with the observed damage. The radiation induced defects in the crystal may be reduced by thermal annealing. Arsenic anti-site defects introduced by irradiation have been shown to anneal at temperatures above 450°C [3]. Detectors were thermally annealed at 450°C in a rapid thermal annealer after exposure to $1 \times 10^{14}\ p\ cm^{-2}$. The electron signal increased as shown in Fig[5] but that due to the holes hardly changed. The response to MIPs, as illustrated in Fig.[6], gave an increase in cce to 50% at a bias of 300 V.

DISCUSSION

The radiation damage measured in GaAs particle detectors correlates very well with calculated values of NIEL for different particle types. The exponential decrease in charge collection efficiency with increasing integrated NIEL cannot easily be explained, as it depends on too many parameters which cannot be measured in isolation. In Si detectors the displacement cross section for protons, neutrons and pions has been found to be almost the same value for each particle at the energy of interest[5]. In the case of GaAs detectors this is not so and a factor of almost 4 has been measured in the damage caused by equivalent fluxes of neutrons and pions, with protons causing roughly three times more damage than neutrons.

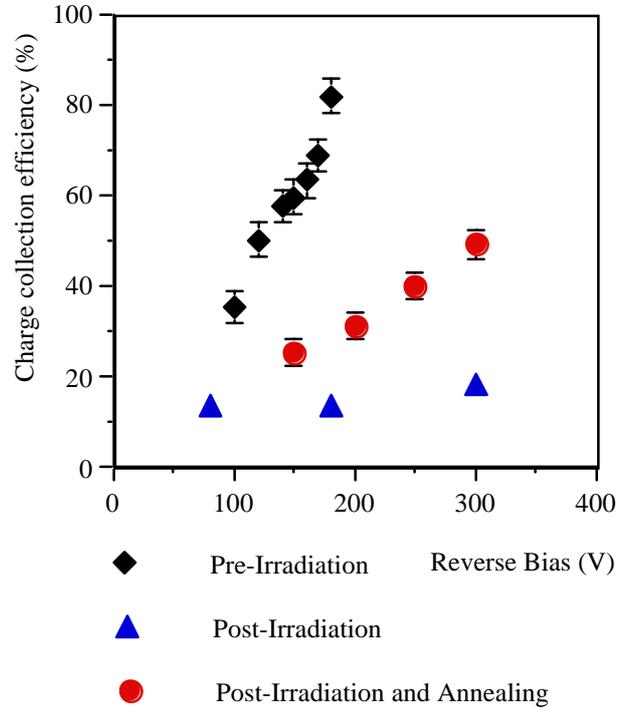

- ◆ Pre-Irradiation
- ▲ Post-Irradiation
- ● Post-Irradiation and Annealing

Fig[6] Improvement in cce after annealing at 450 °C.

| Irradiation type | NIEL Damage factor |
|---|---|
| 1 MeV neutrons | 0.9 |
| 24 GeV/c protons | 2.9 |
| 300 MeV/c pions | 3.6 |

Table 1. Calculated damage factors for different types of radiation by Chilingarov et al[2].

Calculations of the NIEL for each of these particles indicate that the particle equivalence which applies to Silicon detectors is not valid for GaAs based detectors. The GaAs detectors have a greater radiation hardness to neutron irradiation than Silicon but are more susceptible to damage from charged particle irradiation. Indeed, recent studies with Silicon detectors show increased hardness by using harder design rules and subtle processing improvements [6] which should enable them to be used in the harshest areas of the LHC experiments for the duration of the project.

SUMMARY AND CONCLUSIONS

The irradiation studies show that a large signal loss is present in the detectors at high levels of incident charged hadron fluence. This damage is consistent with increased non-ionising energy loss from the charged particle irradiation. Post irradiation annealing helps to recover some of the lost charge by annealing out an electron trap. Further studies are taking place in an attempt to improve the understanding and performance of these devices after irradiation as their present radiation tolerance is not sufficiently hard enough for the forward regions in an LHC experiment.

REFERENCES

[1] S. P. Beaumont et al., GaAs solid state detectors for particle physics.
*Nucl. Instr. and Meth.* **A322** (1992) 472-482.

[2] A. Chilingarov et al., Radiation damage due to NIEL in GaAs particle detectors
Fourth International Workshop on Gallium Arsenide and Related Compounds, Aberfoyle, Scotland, UK, 4-7 June 1996. To be published in *Nucl. Instr. and Meth. A.*

[3] M. Rogalla et al., Radiation studies for GaAs in the ATLAS inner detector.
Fourth International Workshop on Gallium Arsenide and Related Compounds, Aberfoyle, Scotland, UK, 4-7 June 1996. To be published in *Nucl. Instr. and Meth. A.*

[4] R. Worner et al., Electron spin resonance of $As_{Ga}$ antisite defects in fast neutron irradiated GaAs.
*Appl. Phys. Lett.* **40**(2) (1982) 141-143.

[5] G.N. Taylor et al., Radiation induced bulk damage in Silicon detectors.
*Nucl. Instr and Meth. A383 (1996) 144 - 154.*

[6] A. Chilingarov et al., Radiation studies and operational projections for silicon in the ATLAS inner detector.
*Nucl. Instr and Meth. A360 (1995) 432- 437.*